\documentclass[aoas,MSNbibl,nameyear,seceqn,dvips]{arximspdf}
\usepackage{dcolumn,multirow}
\usepackage{graphicx}

\doi{10.1214/14-AOAS774} 
\volume{8}
\issue{4}
\pubyear{2014}
\firstpage{2356}
\lastpage{2377}
\docsubty{FLA}

\makeatletter
\newcolumntype{d}[1]{D{.}{.}{#1}}
\makeatother

\begin{document}
\begin{frontmatter}

\title{Evaluating epoetin dosing strategies using observational
longitudinal data}
\runtitle{Evaluating epoetin dosing strategies}

\begin{aug}
\author[A]{\fnms{Cecilia A.}~\snm{Cotton}\corref{}\ead[label=e1]{ccotton@uwaterloo.ca}\thanksref{T1}}
\and
\author[B]{\fnms{Patrick J.}~\snm{Heagerty}\ead[label=e2]{heagerty@u.washington.edu}\thanksref{T2}}
\runauthor{C.~A. Cotton and P.~J. Heagerty}
\affiliation{University of Waterloo and University of Washington}
\address[A]{Department of Statistics and Actuarial Science\\
University of Waterloo\\
200 University Avenue West\\
Waterloo, Ontario N2L 3G1\\
Canada\\
\printead{e1}}
\address[B]{Department of Biostatistics\\
University of Washington\\
Seattle, Washington 98195\\
USA\\
\printead{e2}} 
\end{aug}
\thankstext{T1}{Supported in part by NIH R01 HL072966 NIH UL1 TR000423.}
\thankstext{T2}{Supported in part by NSERC RGPIN 402474.}

\received{\smonth{5} \syear{2013}}
\revised{\smonth{6} \syear{2014}}

%
\begin{abstract}
Epoetin is commonly used to treat anemia in chronic kidney disease and
End Stage Renal Disease subjects undergoing dialysis, however, there is
considerable uncertainty about what level of hemoglobin or hematocrit
should be targeted in these subjects. In order to address this question,
we treat epoetin dosing guidelines as a type of dynamic
treatment regimen.
Specifically, we present a methodology for comparing the effects of
alternative treatment regimens on survival using observational data.
In randomized trials patients
can be assigned to follow a specific management guideline, but in observational
studies subjects can have treatment paths that appear to be adherent to
multiple regimens at the same time.
We present a cloning strategy in which each subject contributes
follow-up data
to each treatment regimen to which they are continuously
adherent and artificially censored at first nonadherence.
We detail an inverse probability weighted
log-rank test with a valid asymptotic variance estimate that can be used
to test survival distributions under two regimens.
To compare multiple regimens, we
propose several marginal structural Cox proportional hazards models with
robust variance estimation to account for the creation of clones. The
methods are illustrated through simulations and applied to an analysis
comparing epoetin dosing regimens in a cohort of 33,873 adult hemodialysis
patients from the United States Renal Data System.
\end{abstract}

%
\end{frontmatter}

\section{Introduction}\label{ch3introduction}

\subsection{Epoetin treatment for the correction of anemia}\label{introepo}

Erythropoiesis-\break stimulating agents (ESA) are frequently used to correct
for anemia (low red blood cell counts) in patients with a variety of
medical conditions. In particular, recombinant human erythropoietin
(epoetin alfa or, simply, epoetin) has a long history of use in chronic
kidney disease (CKD) and End Stage Renal Disease (ESRD) subjects undergoing
dialysis [\citet{Unger2010}]. Initial evidence supported an
improved quality
of life in subjects whose hemoglobin or hematocrit levels rose after
treatment with epoetin [\citet{CESG1990,Eschbach1994}]. In the
United States
treatment for these patients is covered under Medicare and in 2006 epoetin
was identified as the single largest drug expenditure under
Medicare Part~B [\citet{GAO2006}].

Dialysis subjects are given regular injections of epoetin with the dose
varying over time in response to the subject's changing hemoglobin or
hematocrit levels. Hematocrit is the percentage (\%) of red blood cells
in blood by volume, while hemoglobin (g${}/{}$dl) is a measure of the oxygen
carrying hemoglobin protein found in the blood. Both are used as
measures of anemia and an approximate conversion between the two is to
multiply the hemoglobin measure by three.
Although epoetin has been in widespread use for more
than a decade, there is no consensus as to the optimal hemoglobin or
hematocrit target or dosing algorithm to use in practice. In 2007 the
National Kidney Foundation's Kidney Disease Outcomes Quality
Initiatives (NKF-K/DOQI) panel updated its recommendations for ESA
therapy for anemia in Chronic Kidney Disease to suggest that a target
hemoglobin range of 11.0 to 12.0 g${}/{}$dl (hematocrit 33\% to 36\%) be used
and that hemoglobin targets above 13.0~g${}/{}$dl (hematocrit 39\%) not be
used [\citet{KDOQI2006}].

Several randomized trials have examined the question of what level of
hemoglo\-bin or hematocrit should be targeted in order to improve
quality of life and survival. \citet{Besarab1998} was stopped early
when a higher risk of death and nonfatal myocardial infarction was observed
in dialysis subjects treated to achieve a hematocrit of 42\% versus those
targeted to 30\%. \citet{Singh2006} found an increased risk of a composite
endpoint of several cardiovascular events and death in chronic kidney
disease subjects treated to a target hemoglobin levels of 13.5 g${}/{}$dl versus
11.3 g${}/{}$dl. Around the same time, \citet{Drueke2006} found no significant
difference in all-cause mortality or death from cardiovascular causes
between subjects randomly assigned to have their treatment target a normal
hemoglobin range of 13.0 to 15.0 g${}/{}$dl versus a subnormal range of
10.5 to 11.5 g${}/{}$dl. More recently, \citet{Pfeffer2009} found a
nonsignificant
increased risk of death or nonfatal cardiovascular event in type 2 diabetes
subjects with chronic kidney disease randomized to a target hemoglobin of
13 g${}/{}$dl versus those in a placebo group treated only to maintain a hemoglobin
of about 9.0 g${}/{}$dl. However, there was a significantly higher risk of stroke
and thromboembolic events in the 13 g${}/{}$dl group.

There is also concern that high doses of epoetin may be harmful. Using
a Cox
regression model, \citet{Zhang2004} found that epoetin dose was associated
with increased mortality after adjustment for attained hematocrit level.
\citet{Brookhart2010} found a similar association but only among subjects
with a high achieved hematocrit level. \citet{Zhang2011} found
that among
diabetic patients on dialysis those in the highest epoetin dose group had
a statistically significantly higher risk of experiencing a cardiovascular
event or death. On the other hand, neither \citet{Wang2010} or
\citet{Miskulin2013} found evidence of harm or benefit of higher doses.

Despite the completion of several randomized trials [see additional
references in \citet{Palmer2010}], there still remains
considerable uncertainty
in the best practice for the treatment of CKD-associated anemia. In
particular, the
optimal target hemoglobin/hematocrit range and epoetin dosing algorithm are
unknown [\citet{Unger2010}]. We see this as an opportunity to evaluate
available observational data to determine what evidence such data can provide
regarding epoetin dosing strategies in hemodialysis subjects.

\begin{table}[b]
\tabcolsep=0pt
\caption{Demographic characteristics of 33,873 adult incident
End Stage Renal Disease (ESRD) subjects from United States Renal Data System
(USRDS), 2003}\label{tableUSRDSdemo}
\begin{tabular*}{\tablewidth}{@{\extracolsep{\fill}}@{}ld{2.7}d{2.7}d{2.7}@{}}
\hline
& \multicolumn{1}{c}{\textbf{All subjects}} & \multicolumn{1}{c}{\textbf{Male}} & \multicolumn{1}{c@{}}{\textbf{Females}}\\
\textbf{Characteristic\tabnoteref{tt1}} & \multicolumn{1}{c}{$\bolds{(n=33{,}873)}$} & \multicolumn{1}{c}{$\bolds{(n=17{,}389)}$} & \multicolumn{1}{c@{}}{$\bolds{(n=16{,}484)}$}\\
\hline
Sex \\
Male (\%) & 51.3 & \multicolumn{1}{c}{--} & \multicolumn{1}{c@{}}{--}\\
Age (years) & 66.7~(14.4) & 65.7~(14.8) & 67.7~(14.0)\\
BMI (kg${}/{}$m$^2$) & 27.9~(7.2) & 27.1~(6.3) & 28.8~(8.0) \\
Hematocrit (\%) & 34.4~(9.8) & 34.3~(10.3)& 34.5~(9.3)\\
{Race}\\
\quad White (\%) & 63.2 & 67.1 & 59.2 \\
\quad Black (\%) & 33.4 & 29.5 & 37.6 \\
\quad Other (\%) & 3.4 & 3.5 & 3.3 \\
Comorbid conditions\\
\quad Diabetes (\%) & 63.2 & 60.0 & 66.5 \\
\quad Hypertension (\%) & 83.5 & 82.6 & 84.5 \\
\hline
\end{tabular*}
\tabnotetext{tt1}{Categorical variables are expressed as percentages (\%), continuous
variables are expressed as mean (standard deviation).}
\end{table}

\subsection{United States Renal Data System (USRDS) data set}\label{introUSRDS}

The methods described and developed in this manuscript will be applied
to a
large observational data set from the United States Renal Data System (USRDS).
Data is available on 33,873 adult incident ESRD subjects from the year
2003 with 217,474 total person-months of
observation. The annual death rate is approximately 15\%. Basic
demographic characteristics of the analysis cohort are given in
Table~\ref{tableUSRDSdemo}. For each month, the following information
is available: number of dialysis
sessions reported, number of epoetin doses recorded, total epoetin
dosage (10,000 units), iron supplementation dose, number of days
hospitalized and the
last hematocrit measurement recorded in the month. Dates of baseline
hematocrit measurement, first ESRD service, first transplant and death
are recorded where applicable. Subjects with cancer, human
immunodeficiency virus (HIV) or acquired immunodeficiency syndrome
(AIDS) were excluded form the analysis cohort.

\begin{figure} 

\includegraphics{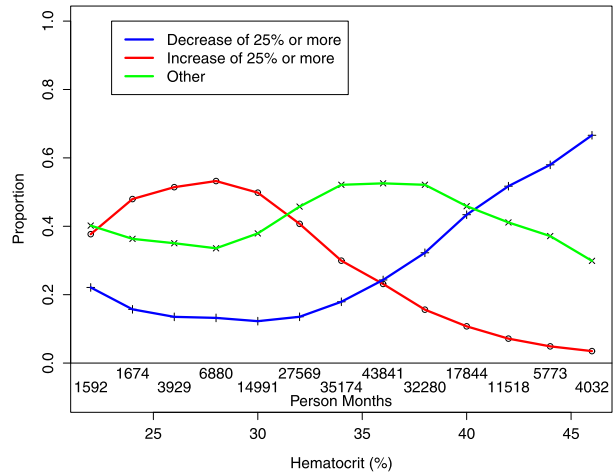}

\caption{Proportion of person-months with 25\% or greater increase or
decrease in epoetin dose by current hematocrit (\%) level.}\label{figepochange}
\end{figure}

Figure~\ref{figepochange} shows how the change in epoetin dose is related
to the current level of hematocrit. Specifically,
we plot the proportion of patient-months where epoetin dose is either
decreased by 25\% or more, increased by 25\% or more, or is maintained
within plus or minus 25\% of the previous month's dose. Here we see the
dynamics of the dose management where subjects with lower hematocrit
levels are most likely to have their epoetin dose increased and those with
high hematocrit are most likely to have their dose decreased. For
patient-months with hematocrit\vadjust{\goodbreak} ranging between approximately 32\% to
40\%, the
most common treatment was to approximately maintain the epoetin
dosage, suggesting that many physicians guiding treatment considered these
to be acceptable hematocrit levels. However, there is still considerable
heterogeneity in treatment changes across all hematocrit levels and we will
exploit this variation to compare outcomes under various epoetin dosing
strategies.

\subsection{Dynamic treatment regimens}

We consider epoetin dosing strategies to be a type of dynamic treatment
regimen. A deterministic dynamic treatment regimen is any sequential decision
strategy, guideline or rule that defines how a subject's current treatment
depends on their measured covariate and possibly treatment histories.
In the case of epoetin dosing, a~treatment regimen constitutes the
target hemoglobin or hematocrit range along with rules that dictate how
the dose of epoetin should be adjusted over time. The specification of candidate
treatment guidelines to be studied is a critical first step in the analysis
process. Once the set of possible guidelines have been defined, one can
retrospectively determine whether or not each subject's treatment was
compliant with a particular regimen, and then base analyses on those months
that were adherent to the regimen under study. Since we wish to characterize
survival under full compliance to a specific dosing guideline,
subjects are typically censored at the first visit when their treatment
trajectory no longer adheres to the regimen under study.

Currently, there are a limited number of statistical methods that permit
direct estimation of the marginal (structural) performance of longitudinal
treatment guidelines, and the evaluation of existing methods is quite limited
with few worked \mbox{examples} and minimal simulation evaluation. An extensive
review of relevant available methods is found in Chapter~5 of
\citet{chakraborty2013}.\vadjust{\goodbreak} Briefly, under appropriate assumptions,
Inverse Probability of Censoring Weights (IPCW) and Marginal Structural Models (MSM)
introduced in \citet{Robins1993}, \citet{Robins1995} and
\citet{Robins2000}
can be used to adjust for measured time-dependent confounding and selection
bias in observational studies. These methods were used in \citet
{Hernan2006}
to compare survival under two dynamic treatment regimens for the initiation
of highly active antiretroviral therapy (HAART) in HIV-infected patients.
Further analyses have compared multiple candidate CD4 cell count thresholds
for the initiation of treatment. For example, \citet{Orellana2010},
\citeauthor{Cain2010} (\citeyear{Cain2010,Cain2011}) have introduced methods
for comparing
multiple regimens by creating an artificial data set in which each subject
contributes observations for each regimen they followed. Recently,
\citet{cotton2011} consider dynamic guidelines and a data augmentation
estimation method, and \citet{shortreed2012} consider quantitative outcomes
relying on the bootstrap for inference. \citet{Robins2008} also
considered a
$g$-estimation approach to finding the optimal regimen, while
\citet{Young2011} focused on analyses using the parametric $g$-formula.

In this paper we focus on the evaluation of treatment guidelines that
target achieving control of a particular intermediate covariate.
The remainder of the article is structured as follows: in
Section~\ref{ch3methods} we introduce notation and adapt the MSM
methods of \citet{cotton2011} and \citet{Orellana2010} to
provide a general
methodology for the comparison of treatment guidelines indexed by a
finite parameter, and that map the observed dose history and intermediate
marker history into a current dose assignment.
In addition, we introduce a new
simple weighted log-rank method to test for differences in the
population survival distribution that would be realized under alternative
dynamic regimens. This test extends ideas in \citet{Pepe1997},
\citet{Pepe1999} and \citet{Zheng2005}. In Section~\ref{ch3sim} we
present a simulation study using a new data generation structure that
permits evaluation of statistical methods for evaluation of dynamic
guidelines where the structural model can be directly determined to
satisfy MSM assumptions. We also evaluate the use of clustered data
sandwich standard errors which are known to be valid but potentially
conservative when used with a MSM with estimated weights. In
Section~\ref{ch3USRDS} we apply the methods to the USRDS
data set of incident hemodialysis subjects. Our case study extends our
previous work [\citet{cotton2011}] and illustrates the methods
using a
relatively long series of longitudinal data that drives adaptive
treatments. As discussed in Section~\ref{introepo}, there is clear
medical motivation to study alternative guidelines in this setting.
Finally, we conclude with a discussion in Section~\ref{ch3discussion}.

\section{Methodology}\label{ch3methods}

\subsection{Notation}

Let $\mathbf{L}_{i}(t)$ be a vector of possibly time-varying covariates
collected on the $i$th subject, $i=1,\ldots,n$, at the $t$th regularly
spaced observation time, $t=0,1,2,\ldots.$ Denote the baseline
covariates by $\mathbf{V}_i = \mathbf{L}_i(0)$. Let $Z_i(t)$ be the
treatment (e.g., a drug dosage) prescribed at visit $t$. It is
assumed that $Z_i(t)$ is determined following the collection of $\mathbf
{L}_i(t)$ and may therefore be influenced by these covariates. Overbars
are used to represent history up to and including time $t$ so that $\bar
{\mathbf{L}}_i(t) = \{ \mathbf{L}_i(s)\dvtx  s=0,\ldots,t \}$. Finally, let
$T_i$ be the event time of interest. Because we are dealing with data
observed at discrete time points, the exact $T_i$ may not be available,
so instead let $D_i(t)$ be the indicator of the event occurring in the
time period $(t,t+1]$. Assume there is no loss to follow-up and the
event is observed in all subjects. Later, this assumption can be
relaxed by using weighting methods similar to those discussed in
Section~\ref{secclones}.

\subsection{Parameterizing the treatment regimen}

We will consider regimens that specify a range of acceptable treatment
values for $Z_i(t)$ given a subject's previous treatment value
$Z_i(t-1)$ and a single time-varying covariate $L_i(t)_1$ [the first
element of the vector $L_i(t)$]. For example, the regimen
%
\begin{equation}
\label{ch3eqn1} Z_i(t)|Z_i(t-1),L_{i}(t)_1,t>0
\in \cases{ Z_i(t-1) \times(p_1, p_2),
\cr
\qquad\mbox{if $L_{i}(t)_1 > b_2$,}
\vspace*{3pt}\cr
Z_i(t-1) \times(p_3, p_4),
\cr
\qquad\mbox{if $L_{i}(t)_1 \in[b_1, b_2]$},
\vspace*{3pt}\cr
Z_i(t-1) \times(p_5, p_6),
\cr
\qquad\mbox{if $L_{i}(t)_1 < b_1$}}
\end{equation}
specifies the range of allowable multiplicative changes
$Z_i(t)$ based on whether $L_{i}(t)_1$ is above, within or below a
target range of $(b_1, b_2)$. This type of regimen is quite flexible
and is fully defined by the set of parameters ($p_1$, $p_2$, $p_3$,
$p_4$, $p_5$, $p_6$, $b_1$, $b_2$). The regimen specification is easily
generalizable to additive changes in dose or dependence on multiple
time-varying covariates from $\bar{\mathbf{L}}_i(t)$.

\subsection{Estimation: Creation of clones}\label{secclones}

For the methods that follow assume that it is unknown which, if any, of
some $K$ treatment regimens of interest a subject was treated under.
Depending on the regimen specifications, it may be possible for a
subject to be adherent to multiple regimens at the same time. In order
to accommodate this, we propose to clone (or replicate) each subject to
create $K$ identical copies of their complete treatment and covariate
history. Let $\mathbf{L}_{ik}(t) = \mathbf{L}_i(t)$ be the copied
vector of time-varying covariates, $\mathbf{V}_{ik} = \mathbf{V}_i$ be
the baseline covariates, $Z_{ik}(t) = Z_i(t)$ be the treatment dose and
let $D_{ik}(t) = D_i(t)$ be the event indicator for subject $i$ under
regimen $k$. Put another way, this refers to subject $i$'s $k$th clone
where $k=1,\ldots,K$.

\begin{figure} 

\includegraphics{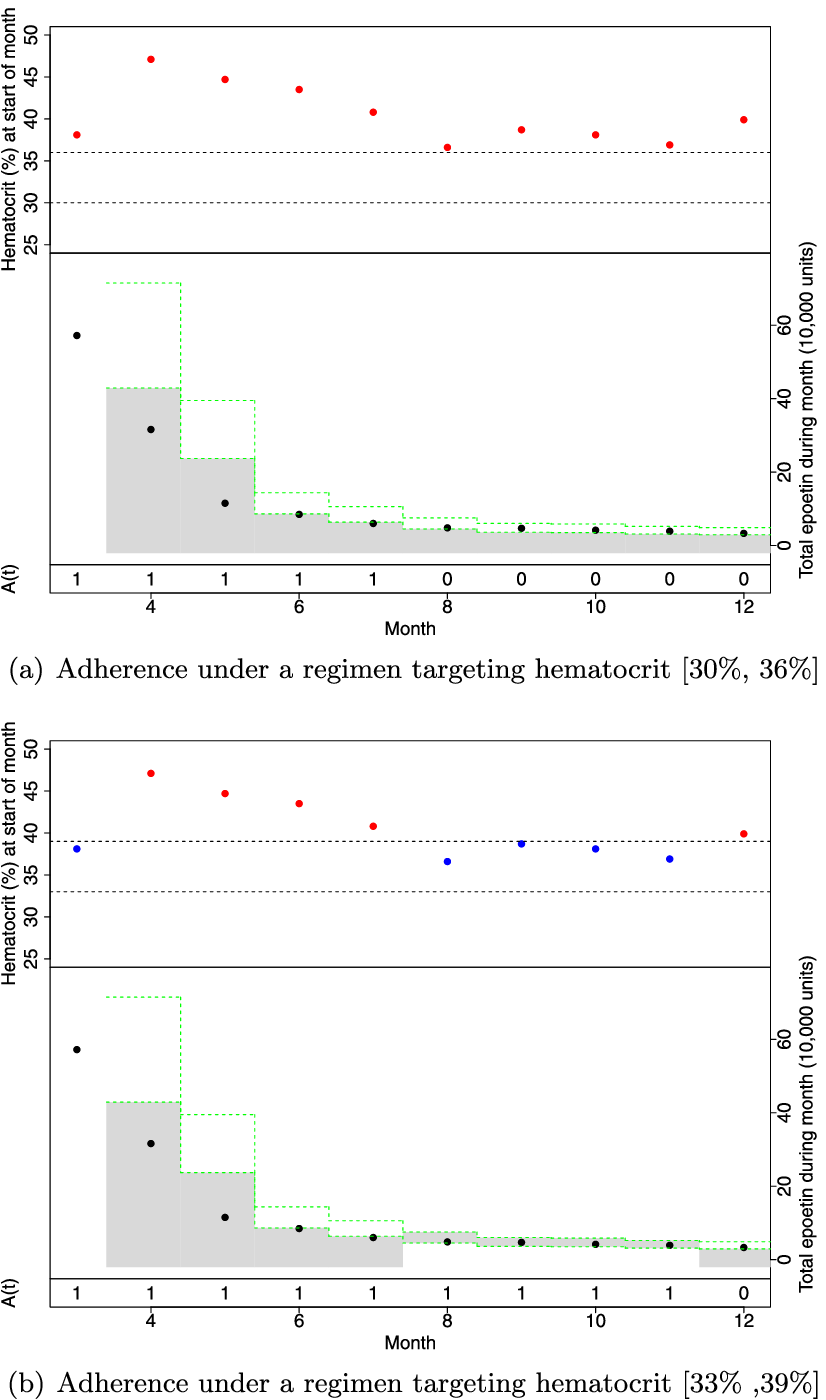}

\caption{Illustration of cloning methodology with hematocrit and
epoetin histories for one subject from the USRDS data set. Upper panels
show monthly hematocrit (\%) and lower panels show total monthly
epoetin dose (10,000 units). In the lower panel the shaded grey regions
indicate where the epoetin dose would have to fall in order to be
compliant with a treatment regimen in the form of equation~(\protect\ref{ch3eqn1}) with $(p_1,p_2,p_3,p_4,p_5,p_6) =
(0,0.75,0.75,1.25,1.25,\infty)$.}\label{figcoadherence}
\end{figure}

Next, we retrospectively determine whether each subject was compliant
with each treatment regimen. We use the terms compliance and adherence
interchangeably and let $A_{ik}(t) = 1$ indicate subject $i$'s
adherence to regimen $k$ at time $t$. Otherwise, $A_{ik}(t) = 0$. Each
clone is artificially censored when they are no longer adherent with
their treatment regimen, and any subject with
zero adherence time to a specific regimen will have fewer than $K$
clones contribute to the analysis.
Let $C_{ik}(t)$ be an indicator of artificial censoring for subject $i$
under regimen $k$ at time $t$. Note that the censoring is fully
determined\vspace*{1pt} through the adherence history $\bar{A}_{ik}(t)$ and given by
$C_{ik}(t) = 1 - I[\bar{A}_{ik}(t) = \bar{1}]$, where $\bar{1}$ is a
vector of ones the same length as $\bar{A}_{ik}(t)$. So $C_{ik}$ is a
vector of zeros followed by ones starting at the first nonadherent
visit. Note that the use of the subscript $k$ on $\mathbf{V}_{ik}$,
$D_{ik}(t)$, $Z_{ik}(t)$ and $\mathbf{L}_{ik}(t)$ is redundant since
the cloning process does not alter the event time or follow-up data, as
it simply defines the artificial censoring time based on adherence.

The key idea is that the creation of clones with appropriate regimen-specific
nonadherence censoring allows us to compare survival under alternative
dosing strategies. Weights are essential to correct for selection bias or
for any factors associated with nonadherence to each specific regimen.
For example, if we only had one regimen of interest, then we would
create only
one parsing of the longitudinal data to reflect observed adherence to the
regimen under study (e.g., not create multiple clones) but would still
need to consider weighting for valid inference regarding survival under
the specific regimen.

If $A_{ik}(t) = A_{il}(t) = 1$, we say that subject $i$ was \emph{coadherent} to regimens $k$ and $l$ at time $t$. The concepts of
cloning and coadherence are illustrated in Figure~\ref{figcoadherence}
with follow-up data from one subject from the USRDS data set. Both
panels display the same observed hematocrit and epoetin dose histories.
The upper and lower panels consider regimens targeting hematocrit
ranges of $[30\%, 36\%]$ and $[33\%, 39\%]$, respectively. Both
regimens specify an allowable multiplicative change in epoetin of plus
or minus 25\% when hematocrit is within the target. For each month, the
grey shaded region indicates the range of epoetin doses that would have
led to compliance with the regimen at that month. From the upper panel
we see that the subject is compliant up to and including month seven.
The lower panel indicates compliance to the higher target regimen up to
month 11. So for months three through seven this subject was coadherent
to both regimens. The subject is artificially censored at months 8
(upper panel) and 12 (lower panel).

The artificial censoring has the potential to induce selection bias.
For example, if subjects with less severe disease, and hence longer
survival times, are less likely to be censored under a particular
regimen, the analysis set will be overrepresented by subjects with less
severe disease. In an unadjusted analysis the effectiveness of the
regimen would be overestimated. We use stabilized inverse probability
weights (IPW) [\citet{Robins1992}] to attempt to adjust for this
potential selection bias:
\[
sw_{ik}(t) = \prod_{s=0}^{t}
\frac{P[C_{ik}(s)=0 | \bar{C}_{ik}(s-1) =
\bar{0}, V_{i}=v_{i}]}{P[C_{ik}(s)=0 | \bar{C}_{ik}(s-1) = \bar{0}, \bar
{L}_{i}(s)=\bar{l}_{i}(s)]}.
\]
At each time point, each clone is weighted by the inverse of
the probability that they remained adherent to their treatment regimen
given their measured covariate history. So adherent clones account for
themselves as well as other similar subjects who were nonadherent to
the regimen and therefore artificially censored. The model in the
numerator includes only baseline covariates and serves to reduce the
variability (i.e., stabilize) of the weights. This occurs because the
probabilities in the numerator and denominator tend to be correlated.
Details of the estimation of these weights have been well covered in
the literature [\citet{Robins2000,Hernan2001}]. Using these
weights creates a pseudo-population in which the probability of
remaining adherent is independent of measured confounders. In order for
these methods to be valid, we must assume that the baseline and
longitudinal information is sufficiently predictive of nonadherence to
satisfy the assumption of effective sequential randomization
[\citet{Hernan2001}]. We discuss assumptions in detail in the next section.

The above cloning process induces a unique correlation structure on the
created clusters of data. Within-clone correlation exists over time due
to the estimation of weights. The weights within clone sets are
expected to be more similar over time than the weights between any two
of a subject's clones. However, between clone correlation also exists
because, if observed (i.e., each clone remained uncensored under their
respective regimen), both clones will have the same time of death.

We have induced a form of censoring and explicitly consider weights to
adjust for this selection bias. However, additional selection bias or
confounding may exist in an observational data set, and additional work
may be needed to conduct valid inference. Additional weights can also
be included to account for censoring due to loss to follow-up or
administrative censoring.

\subsubsection{Causal assumptions}

Suppressing the subject subscript $i$, we assume for each possible
history $\bar{a}$ there is a corresponding counterfactual event time
$T_{\bar{a}}$. In the methods that follow we make the following
assumptions. First, the \emph{sequential randomization} or \emph{no
unmeasured confounders} assumption states that conditional on the
observed covariate and treatment history, the treatment a subject
received at time $t$ is independent of their counterfactual outcomes
$T_{\bar{a}}$:
\[
T_{\bar{a}} \coprod A(k) | \bar{A}(k-1) = \bar{a}(k-1), \bar{L}(k)
= \bar{l}(k), T>u(k),
\]
for all histories $\bar{a}(k-1)$ and $\bar{l}(k)$ where
$u(k)$ is the time of visit $k$ [\citet
{Hernan2001}, \citeauthor{Robins1998} (\citeyear{Robins1998,Robins1999})]. Next, the \emph{positivity}
assumption states that all subjects have a nonzero probability of being
adherent to any regimen:
\[
0 < P \bigl[A(k) = 1 | \bar{A}(k-1) = \bar{a}(k-1), \bar{L}(k) = \bar{l}(k),
T>u(k) \bigr] < 1
\]
with probability 1. Finally, we make the \emph{stable unit
treatment value} assumption that one subject's potential outcome is not
influenced by the treatment allocated to other subjects [\citet
{Rubin1980}].

\subsection{Cloned IPW weighted log-rank test}\label{ch3clonelogrank}

Suppose there are two regimens of interest ($k=1,2$) and each subject
has been cloned as outlined above so that for each subject the survival
of clone $k$ is considered under regimen $k$. Essentially, this is
paired survival data and in order to use a log-rank test, adjustments
must be made for the correlation between pairs/clones. However, the IPW
also needs to be incorporated to adjust for the selection bias induced
by the artificial censoring. The cloned IPW weighted log-rank test
presented here is an extension of the unpaired test described by
\citet{Xie2005} and relies on methods from \citet{Jung1999}
for calculating the standard error of the rank test statistic for
paired survival data. The hypothesis to be tested is that the
cumulative hazard functions are the same under the two regimens:
\begin{eqnarray*}
H_0\dvtx  \Lambda_1(t) &=& \Lambda_2(t)\qquad
\mbox{for all } 0 \le t \le\tau\qquad\mbox{versus}
\\
H_1\dvtx  \Lambda_1(t) &\ne& \Lambda_2(t)\qquad
\mbox{for some } 0 \le t \le\tau,
\end{eqnarray*}
where $\tau$ is the largest time at which both sets of clones have at
least one subject at risk and $\Lambda_1(t)$ and $\Lambda_2(t)$ are the
true underlying cumulative hazard functions.
For subjects $i=1,\ldots,n$ let $(T_{i1}, T_{i2})$ be the i.i.d. paired
(cloned) survival times and $(C_{i1}, C_{i2})$, $i=1,\ldots,n$ be the
i.i.d. cloned censoring times. Then $X_{ik} = \min(T_{ik},C_{ik})$ is the observed event time and $\Delta_{ik} =
I(T_{ik} \le C_{ik})$ is the event indicator for subject $i$ under
regimen $k=1,2$. So the full set of observed data is given by $\{
(X_{i1},X_{i2},\Delta_{i1},\Delta_{i2})$, $i=1,\ldots,n\}$. Note that
our unique cloning correlation structure implies that if $\Delta_{i1} =
\Delta_{i2} = 1$, then $T_{i1} = T_{i2}$.

Using standard survival analysis notation, the event process is given
by $N_{ik}(t) = \Delta_{ik}  I(X_{ik} \le t)$,
and $N_k(t) = \sum_{i=1}^n N_{ik}(t)$ is the total number of deaths
observed under regimen $k$ at or before time $t$. The standard at risk
process is given by $Y_{ik}(t) = I(t \le X_{ik})$, so $Y_k(t) =  \sum_{i=1}^n Y_{ik}(t)$ is the total number of subjects at risk at time $t$
under regimen $k$.

For each regimen assume that the true time-varying subject-specific\break 
weights are $w_{ik}(t)$ and that consistent estimates $\widehat{sw}_{ik}(t)$ are available. Now define a weighted event process
through its derivative as
$dN_k^w(t) = \break\sum_{i=1}^n \,dN_{ik}^w(t)$, where
\[
d N_{ik}^w(t) = w_{ik}(t) \,d N_{ik}(t)
= \cases{ w_{ik}(X_{ik}), &\quad if $t=X_{ik}$ and
$\Delta_{ik}=1$,
\cr
0, &\quad otherwise} %
\]
and a weighted at risk process as
$Y_k^w(t) = \sum_{i=1}^n Y_{ik}^w(t)$ where $Y_{ik}^w(t) = w_{ik}(t)\*
I(X_{ik} \le t)$.
Recall the Nelson estimator of the cumulative hazard and define a
corresponding weighted version:
\[
\hat{\Lambda}_k(t) = \int_0^t
\frac{dN_k(s)}{Y_k(s)}, \qquad \hat{\Lambda}_k^w(t) =
\int_0^t \frac{dN_k^w(s)}{Y_k^w(s)}.
\]

\citet{Jung1999} provides the details of a log-rank test with
correlated survival times. With the addition of time-varying
subject-specific weights, a natural extension of the standard class of
rank statistics is
\[
W^* = \sqrt{n} \int_0^\infty H(t) \bigl[ d\hat{
\Lambda}_1^w(t) - d\hat {\Lambda}_2^w(t)
\bigr]
\]
with
\[
H(t) = \frac{1}{n} \frac{Y_1^w(t) Y_2^w(t)}{Y_1^w(t) + Y_2^w(t)}.
\]

The statistic $W^*$ is equivalent to the usual form of the log-rank
test statistic as the sum over time of the difference in the observed
number of deaths in one group and the expected number of deaths in that
group under $H_0$. For discrete time points $t=1,\ldots,T$ let $d_k(t)
= \sum_{i=1}^n \Delta_{ik} I(X_{ik} = t)$ be the number of deaths
observed in group $k$ at time $t$ and $d_k^w(t) = \sum_{i=1}^n
w_{ik}(t)\Delta_{ik} I(X_{ik} = t)$ be the weighted number of deaths in
group $k$ at time $t$. Then it can be shown that
\[
W^* = \frac{1}{\sqrt{n}} \sum_{t=1}^T
\biggl[ d_1^w(t) - Y_1^w(t)
\biggl( \frac{d_1^w(t) + d_2^w(t)}{Y_1^w(t) + Y_2^w(t)} \biggr) \biggr]. %
\]

In Appendix~A in the supplementary material [\citet{Cotton2014supp}] we derive the above form of $W^*$, show that under
$H_0$, $W^*$ is asymptotically normal with mean 0 and variance $\sigma
^2$, and give a consistent estimator for $\sigma^2$.

\subsection{Cloned marginal structural Cox proportional hazards models}
\label{ch3cloneMSM}

\subsubsection{Comparison of two treatment regimens}

The usual Cox proportional hazards adherence-based MSM [\citeauthor{Hernan2000} (\citeyear{Hernan2000,Hernan2001}),
\citet{RobinsFinkelstein2000}] can be used with the
cloned survival data
provided that valid standard errors are used to account for the ``clone
clusters.'' In the simplest setting of comparing two treatment regimens
with known regimen membership at baseline, we can specify a
proportional hazards marginal association model:
\[
\lambda_T(t|G_i,\mathbf{V}_i) =
\lambda_0(t) \exp \bigl(\beta_1 G_i +
\bolds\alpha' \mathbf{V}_i \bigr),
\]
where $\lambda_0(t)$ is an unspecified baseline hazard
function, $G_i$ is an indicator of regimen assignment and $V_i$ is a
set of baseline (nontime-varying) covariates. If there were no
censoring/nonadherence and regimen membership $G_i$ had been randomly
assigned, then there would be no confounding and the parameter $\beta
_1$ would have a causal interpretation.

Specifically, $\exp(\beta_1)$ is the causal hazard ratio comparing the
two regimens. Most standard statistical software packages do not allow for
the inclusion of subject-specific time-varying weights in fitting a
Cox model. However, the model can be fit using weighted pooled logistic
regression weighted by $\widehat{sw}_i(t)$ with each subject visit treated
as a single observation:
\[
\operatorname{logit} P \bigl[D_i(t)=1|D_i(t-1)=0,G_i,
\mathbf{V}_i \bigr] = \beta_0(t) + \beta_1
G_i + \bolds\alpha' \mathbf{V}_i.
\]

Here $\beta_0(t)$ is a time-specific intercept usually fit as
a spline. While this yields a consistent estimate of $\exp(\beta_1)$,
the estimated standard error may be invalid since the estimation of the
weights induces a within-subject correlation. In order to overcome
this, the model is fit using a Generalized Estimating Equations (GEE)
approach with working independence [\citet{LiangZeger1986}].

For the cloned data setting we proceed in the same manner but treat
each of the $2n$ clones as independent observations. Let
$G_{ik}=I[k=2]$ be the indicator that the clone is followed under
regimen 2. To fit the MSM, each clone visit is now treated as a single
observation in the logistic model:
\[
\operatorname{logit} P \bigl[D_{ik}(t)=1|D_{ik}(t-1)=0,G_{ik},
\mathbf{V}_{ik} \bigr] = \beta_0(t) + \beta_1
G_{ik} + \bolds\alpha' \mathbf{V}_{i}.
\]

We assume working independence, but based on results in \citet
{Lee1992} for the Cox model, the estimated regression parameters will
still be consistent.
A consistent variance estimate can be obtained from the standard GEE
sandwich covariance estimate if weights are known, and will provide
conservative standard errors with estimated weights [\citet{Hernan2001}].

\subsubsection{Extension to multiple treatment regimens}\label{ch3models}

Instead of comparing just two treatment regimens, suppose there are
multiple regimens to be compared simultaneously. Let $G_{ik}=k$,
$k=1,2,\ldots,K$ indicate a clone's treatment regimen assignment. There
are a variety of different Cox proportional hazard MSMs that can be
considered. In general, let
%
\begin{equation}
\label{cloneMSM} \lambda_T(t|G_{ik},\mathbf{V}_{i})
= \lambda_0(t) \exp \bigl[\beta(G_{ik},t) + \bolds
\alpha' \mathbf{V}_{i} \bigr],
\end{equation}
where $\beta(G_{ik},t)$ is a smooth function of both the
observation time $t$ and the regimen number $G_{ik}$. Special cases of
the above model include assuming a linear regimen effect $ \beta
(G_{ik},t) = \beta G_{ik} $ or treating regimen as a factor variable
with or without interactions with time. A more flexible model would
include splines in regimen number and/or the effect of time.
The interpretation of $\beta$ will be as a causal hazard ratio,
although the precise interpretation will depend on the model specification.
We assume that the effect of the covariates $\mathbf{V}$ is constant
across the comparison of any two regimens at any times.

\section{Simulation study}\label{ch3sim}

A simulation study was undertaken to: (1) illustrate the structural
models described above in a setting where counterfactual outcomes
satisfy known relationships, (2) evaluate the performance of the point
estimation strategy, and (3) evaluate the performance of the sandwich
standard errors. The first and third of these goals have not been fully
addressed in the existing literature.

For each of $K=6$ regimens we simulated $n_k=2500$ survival times using
an exponential distribution with rate parameter $\lambda_k$. This
generated continuous simulated survival times $T_i$ for $i=1,\ldots,
n=\sum n_k={}$15,000 subjects each under full adherence to one regimen. We
discretize $T_i$ to $D_i(t) = I[t<T_i \le t+1]$, the indicator of death
in the next time period. Let $\tilde{k}$ represent the regimen under
which subject $i$'s survival time was generated. The adherence
indicators for regimen $\tilde{k}$ are $A_{i\tilde{k}}(t) =1$ for
$t=0,\ldots,\operatorname{int}(T_i)$, where $\operatorname{int}(T_i)$
is the largest integer less than~$T_i$.

Each simulated subject is cloned and their adherence to the $K-1$ other
regimens is simulated based on a set of fixed coadherence
probabilities.The rate parameters $\lambda_k$ are selected in such a
way that the hazard ratios $\lambda_k/\lambda_3$ for $k=2,4,5$ are the
same for the original raw data and the cloned data. Details are given
in Appendix~B.1 in the supplementary material [\citet{Cotton2014supp}]. The data generation method does not guarantee that
the hazard ratios $\lambda_1/\lambda_3$ and $\lambda_6/\lambda_3$ are
the same before and after the cloning, so the results for these two
regimens are not included.
To induce selection bias through artificial censoring, a scalar
baseline covariate $V_i$ associated with both survival time and
coadherence (and therefore censoring) is included. We consider three
levels of selection bias: none, moderate and severe. For details, see
Appendix B.2.

The concept of coadherence in the simulation study relates directly to
the comparison of multiple treatment regimens. Consider two regimens of
the form of equation~(\ref{ch3eqn1}) with overlapping target ranges.
At any given month a subject is much more likely to be adherent to both
these regimens than they would be to be adherent to two regimens with
nonoverlapping target ranges. In addition, any baseline covariate
that's used in the decision of how to change treatment in response to
changing hematocrit may be associated with coadherence of two regimens.

The aggregated results of 500 simulations are presented in Table~\ref
{ch3table2}. Regimen $k=3$ is considered the reference regimen. The
first two columns of the table present the true underlying hazard
ratios $\lambda_k/\lambda_3$ for $k=2,4,5$ and the median of the 500
estimated hazard ratios based on the original $n$ simulated fully
compliant event times. The agreement between these two columns
demonstrates that the true underlying hazard ratios can be estimated
from the data despite the discretization of the event times.

\begin{table}
\tabcolsep=0pt
\caption{Estimated median hazard ratios (HR), empirical standard errors
(ESE), average standard errors (ASE) and empirical 95\% confidence
interval coverages (ECP) from cloning methodology simulations at three
levels of induced selection bias (500 replications each with $n_k=2500$)}\label{ch3table2}
\begin{tabular*}{\tablewidth}{@{\extracolsep{\fill}}@{}lcccccccccc@{}}
\hline
& \multirow{3}{19pt}{\centering{\textbf{True HR}}} & \multirow{3}{37pt}{\centering{\textbf{Complete data HR}}} & \multicolumn{4}{c@{}}{\textbf{Clones, unweighted}} & \multicolumn{4}{c@{}}{\textbf{Clones, IPW weighted}}\\[-6pt]
& & & \multicolumn{4}{c@{}}{\hrulefill} & \multicolumn{4}{c@{}}{\hrulefill}\\
\textbf{Regimen} &  & & \textbf{HR} & \textbf{ESE} & \textbf{ASE} & \textbf{ECP} & \textbf{HR} & \textbf{ESE} & \textbf{ASE} & \textbf{ECP}\\
\hline
\multicolumn{11}{@{}c@{}}{\textit{No induced selection bias}}\\
2 vs 3 & 0.91 & 0.90 & 0.89 & 0.0204 & 0.0233 & 92.8 & 0.89 & 0.0221 & 0.0232 & 90.6 \\
4 vs 3 & 1.17 & 1.17 & 1.18 & 0.0305 & 0.0316 & 94.8 & 1.18 & 0.0330 & 0.0315 & 92.8 \\
5 vs 3 & 1.28 & 1.29 & 1.31 & 0.0341 & 0.0352 & 90.2 & 1.31 & 0.0352 & 0.0357 & 88.2
\\[3pt]
\multicolumn{11}{@{}c@{}}{\textit{Moderate selection bias}}\\
2 vs 3 & 0.91 & 0.90 & 0.89 & 0.0206 & 0.0230 & 94.2 & 0.89 & 0.0213 & 0.0225 & 92.4 \\
4 vs 3 & 1.17 & 1.18 & 1.27 & 0.0308 & 0.0340 & 11.8 & 1.17 & 0.0301 & 0.0314 & 94.8 \\
5 vs 3 & 1.28 & 1.29 & 1.40 & 0.0357 & 0.0376 & \phantom{0}9.2 & 1.30 & 0.0346 & 0.0355 & 92.8
\\[3pt]
\multicolumn{11}{@{}c@{}}{\textit{Severe selection bias}}\\
2 vs 3 & 0.91 & 0.91 & 0.90 & 0.0217 & 0.0227 & 94.0 & 0.89 & 0.0217 & 0.0218 & 92.6 \\
4 vs 3 & 1.17 & 1.18 & 1.37 & 0.0335 & 0.0365 & \phantom{0}0.0 & 1.17 & 0.0302 & 0.0314 & 96.4 \\
5 vs 3 & 1.28 & 1.29 & 1.49 & 0.0416 & 0.0401 & \phantom{0}0.0 & 1.30 & 0.0362 & 0.0354 & 90.2 \\
\hline
\end{tabular*}
\end{table}

The next four columns present results from the cloned, unweighted data
and demonstrate the effect of the selection bias induced through the
coadherence probabilities described in Appendix B.2. In all scenarios
the data was generated without selective nonadherence between regimens
2 and 3 and the median estimated hazard ratio and the coverage of the
95\% confidence intervals just below the nominal level. The coadherence
probabilities do induce substantial selection bias between regimens 4
and 3 and regimens 5 and 3. In the severe selection bias scenario the
empirical coverage probability of the confidence intervals is zero.

The final four columns of Table~\ref{ch3table2} give the results of
weighting the clones by the estimated IPW. A logistic model for
adherence given the $V_i$ covariate is used for the weights. In all
three scenarios the median estimated hazard ratios are very close to
the truth. In the moderate and severe selection bias scenarios the
coverage of the confidence intervals is greatly improved, although it
is slightly below the nominal level in several cases.
In cases without selection bias, we suspect that the lower than nominal
coverage rates in the IPW analysis are due to instability of
inefficiency of the IPW estimates after the inclusion of unnecessary weights.
The empirical and average standard errors for the unweighted and
weighted clones are comparable.
This supports the claim that the zero coverage in the unweighted case
is due to bias in the estimates as opposed to underestimating the variance.

\section{Application to USRDS data set}\label{ch3USRDS}

In this section we apply the cloning methodology to a large data set
from the USRDS introduced in Section~\ref{introUSRDS}. Analysis begins
at month 3 (since the initial dosing strategy is different from the
maintenance dosing strategy that we wish to study), with up to 9 months
of follow-up data per subject ($t=0,\ldots,9$).

\subsection{Naive analyses}

First we use Cox proportional hazards regression to conduct naive
analyses of
the acute association between mortality and epoetin dose. The last month's
assigned dose is treated as a time-dependent covariate in models
adjusted for baseline covariates (age, sex, race, diabetes and
hypertension). Models are fit both with and without time-dependent
hematocrit (measured as the average of the last two months' values).
Both models yield essentially the same result. The estimated log hazard
ratio associated with one unit increase in epoetin dose is 0.031
(0.028, 0.034), suggesting that higher epoetin doses are associated
with an increased risk of mortality, as one would expect higher
hematocrit levels are associated with a reduced risk of mortality [log
hazard ratio of $-$0.035 ($-$0.032, $-$0.038)]. The naive analysis is
difficult to translate into recommendations for epoetin dosing
strategies since trying to attain a high hematocrit level and a low
epoetin dose may be incompatible in most patients.

\subsection{Dynamic treatment regimens for epoetin dosing}

We consider multiple dynamic treatment regimens for epoetin dose
$Z_i(t)$ given current hematocrit level $L_{i}(t)_1$ and previous dose
$Z_i(t-1)$ of the form in equation~(\ref{ch3eqn1}), where
\[
Z_i(t)|Z_i(t-1),L_{i}(t)_1,t>0
\in\cases{
Z_i(t-1) \times \bigl(-\infty, -(1-p) \bigr),
\cr
\qquad\mbox{if $L_{i}(t)_1 > x-3$},
\vspace*{3pt}\cr
Z_i(t-1) \times(0.75, 1.25),
\cr
\qquad\mbox{if $L_{i}(t)_1 \in[x-3, x+3]$,}
\vspace*{3pt}\cr
Z_i(t-1) \times \bigl((1+p), \infty \bigr),
\cr
\qquad\mbox{if $L_{i}(t)_1 < x+3$,}}
\]
with $x$ representing the midpoint of the target hematocrit range of
$(x-3, x+3)$ and $p$ controlling the allowable multiplicative change in
dose outside the target range. We consider regiments with $x=31,\ldots,
40$ and $p=0.05,\ldots,0.50$ in $0.05$ increments and refer to the
regimens by the notation $\mathcal{G}(p,x-3,x+3)$. The regimen $\mathcal
{G}(0.25,30,36)$ is used as the baseline regimen for comparison
purposes. A~target range of $[30\%,36\%]$ was selected to mimic the
subnormal targets used in several of the clinical trials referenced in
Section~\ref{introepo}.

The logistic model for the denominator of the IPW includes a spline in
time along with the baseline covariates gender, age, race and
indicators of diabetes and hypertension and the time-varying covariates
previous month's total epoetin dose and indicators of whether the
average of the current and previous hematocrit was in the ranges
$(0,28]$, $(28,32]$, $(36,40]$, $(40,\infty)$, as well as the
difference between the current and the previous hematocrit. The model
for the numerator was the same as above but did not include the
time-varying covariates. Additional weights were calculated for
administrative censoring (i.e., clones who were still alive and
compliant to their regimen at month 12) and censoring due to loss to
follow-up (i.e., clones who were alive and compliant at a final
recorded visit occurred prior to month 12). All weight models were
stratified by regimen. The three stabilized weights were multiplied to
give the final weight used in the analyses. The weight specification is
key for valid causal inference.
All available potential confounders, in particular, key variables known
to drive changes in epoetin dosing and epoetin response [for a summary
see Table~1 of \citet{Miskulin2009}], were included in the model
to guard as best as possible against potential violation of the \emph
{no unmeasured confounders} assumption. However, we did not have access
to detailed lab data nor do we have clinic data for additional
adjustment and we acknowledge these potential limitations.

\subsection{Results of cloned IPW weighted log-rank test}

The cloned IPW\break weighted log-rank test from Section~\ref
{ch3clonelogrank} was applied testing the equivalence of survival
under various regimens with $\mathcal{G}(0.25,30,36)$. The results of a
subset of the tests are given in Table~\ref{ch3logranktable}. We
reject the null hypothesis that the two survivor functions are equal
($p<0.05$) in all cases except the comparisons of $\mathcal
{G}(0.10,30,36)$ and $\mathcal{G}(0.40,30,36)$ [the two regimens that
share a common hematocrit target range with the reference regimen
$\mathcal{G}(0.25,30,36)$] and $\mathcal{G}(0.40,32,38)$ (a regimen
with a~slightly higher target range and requiring more aggressive
epoetin dose changes). The results of these tests indicate that there
are significant differences in survival across possible epoetin dosing
regimens. We will proceed with a regression analysis to try and capture
trends in the survival across regimens.

\begin{table} 
\tabcolsep=0pt
\tablewidth=250pt
\caption{Results of cloned IPW weighted log-rank test, USRDS data,
$p$-values for tests of regimens $\mathcal{G}(p,x-3,x+3)$ versus
regimen $\mathcal{G}(0.25,30,36)$}\label{ch3logranktable}
\begin{tabular*}{\tablewidth}{@{\extracolsep{\fill}}@{}lccc@{}}
\hline
&\multicolumn{3}{c@{}}{$\bolds{p}$}\\[-6pt]
&\multicolumn{3}{c@{}}{\hrulefill}\\
& \textbf{0.10} & \textbf{0.25} & \textbf{0.40}\\
\hline
31 & \phantom{$<$}0.032 & \phantom{$<$}0.001 & \phantom{$<$}0.001 \\
33 & \phantom{$<$}0.114 & \multicolumn{1}{c}{Reference} & \phantom{$<$}0.423 \\
$x$ 35 & $<$0.001 & \phantom{$<$}0.006 & \phantom{$<$}0.077 \\
37 & $<$0.001 & $<$0.001 & $<$0.001\\
39 & $<$0.001 & $<$0.001 & \phantom{$<$}0.001\\
\hline
\end{tabular*}
\end{table}

\subsection{Cloned MSM results}

\begin{figure}

\includegraphics{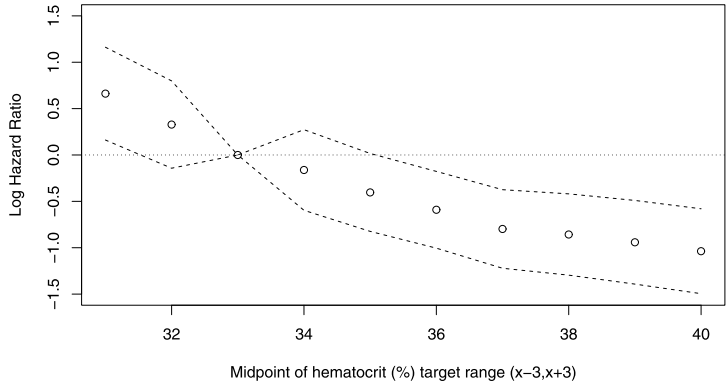}

\caption{Estimated log hazard ratios at 9 months for regimens $\mathcal
{G}(0.25,x-3,x+3)$ versus regimen $\mathcal{G}(0.25,30,36)$, cloned Cox
marginal structural model (MSM), regimen treated as a factor variable
with a log time interaction, USRDS data.}
\label{ch3fig4}
\end{figure}

Several models of the form of equation~(\ref{cloneMSM}) have been fit
to the data. The comparison of target hematocrit ranges is of primary
interest, so the focus in on comparison of the regimens $\mathcal
{G}(0.25,x-3,x+3)$. In models where regimen number is considered as a
linear variable, the regimens are ordered by increasing target range
midpoint. In models where regimen number is treated as a factor
variable, pairwise comparisons between the regimen $\mathcal
{G}(0.25,30,36)$ and the nine other regimens with $p=0.25$ are considered.
\begin{longlist}[2.]
\item[1. \textit{Linear regimen effect}.]
The estimated causal log hazard ratio for a one unit increase in
regimen number is $-$0.017 ($-$0.022, $-$0.011). This is equivalent to a
causal hazard ratio of 0.983 (0.978, 0.989), suggesting that regimens
that target higher hematocrit ranges provide a small but statistically
significant reduction in mortality.

\item[2. \textit{Regimen treated as a factor variable}.]
The estimated causal hazard ratios comparing each treatment regimen
with $\mathcal{G}(0.25,30,36)$ are presented in Table~\ref{ch3table3}.
These results also suggest that regimens targeting a higher hematocrit
range yield a higher rate of survival. The hazard ratios are
significantly different from one another for regimens targeting a range
of $[34\%, 40\%]$ or higher.

\item[3. \textit{Linear regimen effect with a log time interaction}.]
Table~\ref{ch3table4} gives the fitted causal hazard ratios for a one
unit increase in regimen number at each observation month. For all
months after month 3 (the baseline month) the hazard ratio is
statistically significantly less than zero, suggesting that survival
improves as the hematocrit target range midpoint increases. There is a
trend in decreasing hazard ratio over time, suggesting that at later
months there is an increased effect of treating with a regimen
targeting a higher range.

\begin{table} 
\tabcolsep=0pt
\tablewidth=250pt
\caption{Estimated hazard ratios from a cloned Cox marginal structural
model (MSM), regimen treated as a factor variable with reference
$\mathcal{G}(0.25,30,36)$, USRDS data}\label{ch3table3}
\begin{tabular*}{\tablewidth}{@{\extracolsep{\fill}}@{}lcc@{}}
\hline
\textbf{Regimen comparison} & \textbf{Hazard ratio} & \textbf{95\% CI} \\
\hline
$\mathcal{G}(0.25,28,34)$ vs $\mathcal{G}(0.25,30,36)$ & 1.056 & (0.985, 1.132) \\
$\mathcal{G}(0.25,29,35)$ vs $\mathcal{G}(0.25,30,36)$ & 1.027 & (0.959, 1.101) \\
$\mathcal{G}(0.25,30,36)$ vs $\mathcal{G}(0.25,30,36)$ & \multicolumn{1}{c}{Reference} & --\\
$\mathcal{G}(0.25,31,37)$ vs $\mathcal{G}(0.25,30,36)$ & 0.984 & (0.919, 1.054) \\
$\mathcal{G}(0.25,32,38)$ vs $\mathcal{G}(0.25,30,36)$ & 0.958 & (0.895, 1.026) \\
$\mathcal{G}(0.25,33,39)$ vs $\mathcal{G}(0.25,30,36)$ & 0.940 & (0.878, 1.006) \\
$\mathcal{G}(0.25,34,40)$ vs $\mathcal{G}(0.25,30,36)$ & 0.920 & (0.859, 0.985) \\
$\mathcal{G}(0.25,35,41)$ vs $\mathcal{G}(0.25,30,36)$ & 0.913 & (0.853, 0.977) \\
$\mathcal{G}(0.25,36,42)$ vs $\mathcal{G}(0.25,30,36)$ & 0.915 & (0.854, 0.980) \\
$\mathcal{G}(0.25,37,43)$ vs $\mathcal{G}(0.25,30,36)$ & 0.915 & (0.855, 0.980) \\
\hline
\end{tabular*}
\end{table}

\begin{table}[b] 
\tabcolsep=0pt
\tablewidth=250pt
\caption{Estimated hazard ratios by month for a one unit increase in
the midpoint of the target hematocrit range $x$ in regimens $\mathcal
{G}(0.25,x-3,x+3)$, cloned Cox marginal structural model (MSM), linear
regimen effect with a log time interaction, USRDS data}\label{ch3table4}
\begin{tabular*}{\tablewidth}{@{\extracolsep{\fill}}@{}lcc@{}}
\hline
\textbf{Month} & \textbf{Hazard ratio} & \textbf{95\% CI} \\
\hline
\phantom{0}3 & 0.999 & (0.993, 1.004) \\
\phantom{0}4 & 0.950 & (0.940, 0.959) \\
\phantom{0}5 & 0.914 & (0.898, 0.930) \\
\phantom{0}6 & 0.885 & (0.864, 0.907) \\
\phantom{0}7 & 0.862 & (0.837, 0.887) \\
\phantom{0}8 & 0.842 & (0.814, 0.871) \\
\phantom{0}9 & 0.825 & (0.794, 0.857) \\
10 & 0.810 & (0.777, 0.845) \\
11 & 0.797 & (0.761, 0.834) \\
12 & 0.785 & (0.747, 0.824) \\
\hline
\end{tabular*}
\end{table}

\item[4. \textit{Regimen treated as a factor variable with a log time interaction}.]
Recall that in this model within each regimen pair the log hazard ratio
is assumed to be linear in $\log t$. A plot of the estimated causal log
hazard ratios and pointwise 95\% confidence intervals at month 9 is
given in Figure~\ref{ch3fig4}. Informally, in both graphs an initial
downward trend in estimates is seen for regimens with higher target
ranges followed by possibly a flatter trend at the highest target
ranges. The causal log hazard ratio for full compliance to the regimen
$\mathcal{G}(0.25,34,40)$ to survival under full compliance to the
regimen $\mathcal{G}(0.25,30,36)$ was $-$0.51 $(-0.77, -0.25)$ at month
6 and $-$0.80 $(-1.22, -0.37)$ at month 9.
\end{longlist}

These four models all show the same trend in survival when considering
the regimens $\mathcal{G}(0.25,x-3,x+3)$, $x=31,\ldots,40$ as defined
above. In general, regimens with target ranges above the reference
range of $[30\%, 36\%]$ provide a survival advantage when compared to
$\mathcal{G}(0.25,30,36)$. These gains persisted with the inclusion of
log(month) in the model. It is apparent in Figure~\ref{ch3fig4} that
while regimens with a higher target range yield a survival advantage,
the improvement remains relatively constant for regimens with targets
at or above $[34\%, 40\%]$.

Similar models to those above were fit to the data for regimens,
allowing varying multiplicative changes in epoetin doses outside the
target range. The specific results are not reported here, but in
general there were small (statistically insignificant) survival
advantages for regimens with smaller $p$, that is, those that allowed
smaller changes in epoetin dose when hematocrit was outside the $[30\%,
36\%]$ target range.

\section{Discussion}\label{ch3discussion}

In this article we detailed a cloning methodology for comparing dynamic
treatment regimens when regimen membership is not known at baseline. Our
goal was to perform an appropriate analysis of motivating data from
dialysis patients. In order to implement cloning methods, we first detailed
a simple log-rank test, and then proposed use of clustered survival methods
for regression inference. In order to evaluate the proposed methods, we
provide a novel simulation scenario that can control the structural parameters.
The methods are based on replicating or cloning each subject and
considering the adherence of each clone's treatment history to a
particular treatment regimen under consideration. Clones are artificially
censored at their first nonadherent observation and IPW are used to adjust
for the induced selection bias. If there are only two treatment regimens
under comparison, we have shown that a cloned IPW weighted log-rank test
can be used to test for equality of the survivor functions. The proposed
variance estimator appropriately adjusts for the correlation within clones.
When multiple treatment regimens are under consideration a Cox proportional
hazards adherence-based MSM can be used to compare survival under the
regimens. The structural regression model can take a variety of forms. In
particular, one can choose to model regimen number as a linear or factor
variable and choose whether or not to include interactions with time. In
all cases, a consistent estimate of the causal hazard ratio is available.

For epoetin dosing in incident ESRD hemodialysis subjects, we have
applied this methodology to a large USRDS data set to compare survival
across multiple treatment regimens. As a demonstration, a variety of
models were fit, but all essentially gave the same conclusion. Subjects
tend to experience lower all-cause mortality when treated under epoetin
dosing rules with higher hematocrit target ranges. However, there is
evidence that there is no further gain in survival under regimens with
targets above $[34\%, 40\%]$. This result is scientifically meaningful,
especially in light of the uncertainty in best practice for the
treatment of CKD/ESRD-associated anemia.

This methodology is appealing because there is no requirement that regimens
under consideration be of the same form. In fact, as long as adherence
can be precisely determined, the treatment regimens can be extremely
complex and depend on multiple time-varying covariates or prognostic
factors. Through cloning, all subjects contribute information to all
regimens to which they were continuously adherent. However, due to
artificial censoring, any follow-up after a nonadherent visit is discarded.
Further work is warranted to explore methods that might overcome or relax
this requirement.

The current methods do not explicitly distinguish between different
types of adherence (above, within or below target). A possible
extension would be to include patient status relative to the target in
the adherence model. Alternatively, it would be possible to consider
adherence as a multinomial variable and simultaneously model the
different types of nonadherence, for example, nonadherence due to
insufficient increase in dose when the subject is below target,
insufficient dose decrease when above target, unnecessary dose increase
when within target or unnecessary dose decrease when within target.
This would complicate the definition of the stabilized weights but
warrants further investigation.

\section*{Acknowledgments}
We gratefully acknowledge our anonymous referees and Associate Editor
for helpful comments and suggestions on an earlier draft of this manuscript.


\begin{supplement}[id=suppA]
\stitle{Appendices}
\slink[doi]{10.1214/14-AOAS774SUPP} 
\sdatatype{.pdf}
\sfilename{AOAS774\_supp.pdf}
\sdescription{The supplementary material includes Appendix A:
Asymptotics of Cloned IPW Weighted Log-Rank Test and Appendix B:
Simulation Details.}
\end{supplement}


%

\printaddresses
\end{document}